# New generation B-Field and RAD-tolerant DC/DC power converter for on-detector operation

**A. Lanza,**[a] **E. Romano**[a,b] **and S. Selmi**[c]

[a] *Istituto Nazionale di Fisica Nucleare (INFN), sezione di Pavia,*
   *Via Bassi 6, 27100 Pavia (PV), Italy*

[b] *Università degli Studi di Pavia,*
   *Strada Nuova 65, 27100 Pavia (PV), Italy*

[c] *CAEN S.p.A.,*
   *Via della Vetraia 11, 55049 Viareggio (LU), Italy*
   *E-mail*: emanuele.romano@unipv.it

ABSTRACT: The increase in the number of readout channels in new detectors, like the Micro Pattern Gas Detectors (MPGD), in the order of several millions, requires a large amount of electrical power to supply the front-end electronics, up to hundreds kW. If this power is generated at long distances from the detector, the voltage drop on the connection cables puts serious constraints to the supply current, to the wire cross-section and to the power distribution. A large amount of voltage drop on the cables, apart an increased power dissipation on wire resistance, determines regulation issues on the load in case of current transients. To mitigate these problems, a new generation DC/DC converter, working in a heavily hostile environment and with a power density greater than 200 W/dm$^3$, was developed. It is modular, with up to four independent modules, eight channels each, collected in a water-cooled crate, and can supply the load with an adjustable 10 to 12 V output up to 170 W per channel. In this contribution, the design constraints of such a converter are analysed, taking as a basis the environmental, electrical and mechanical requirements of the ATLAS New Small Wheel (NSW) project. Thermal considerations require the converter to be water-cooled, and the dimensional constraints impose the adoption of an innovative design to convey the dissipated heat towards the heat exchanger. The control and monitoring system allows for the full remote management of the converter. Main electrical parameters were measured and are reported. The converter was also characterized in a harsh working environment, with radiation tests in the CERN CHARM facility beyond the limits estimated for ten years operation in ATLAS, and with magnetic field tests in various orientations, using different magnets at CERN up to 1.3 T. A better common mode noise filtering design improvement was implemented after the first characterization. Final performance measurements after the modifications are also reported.

# 1. Introduction

Most of the recent upgrades in large physics experiments use new types of detectors, like the Micro Pattern Gaseous Detectors (MPGD). They have increased number of readout channels and front-end electronics boards, integrating state-of-the-art FPGAs. This tendency requires upgrading the power distribution strategy and the need of managing a larger number of independent channels, distributing the power with high granularity and requiring the delivery of low supply voltages (up to few V) with currents up to few A. In comparison to the standard power distribution strategy (where the converters for gaseous detectors are located on the balconies of the experiments), it is not anymore possible to place the power supplies too far from the relative detector. Lower voltages with higher current result in higher voltage drop and the need for a higher cable cross section. Moreover, the voltage drop on the cables, apart from a waste of power, would introduce problems in the precise regulation of the voltages on the loads. The new power distribution strategy (Figure 1) reconceives the multi-stage conversion without adding points of load (the function of the rack in the balcony is moved and integrated in the detector) and provide remotely the primary supply of 300 V (simplifying the design of the primary converter).

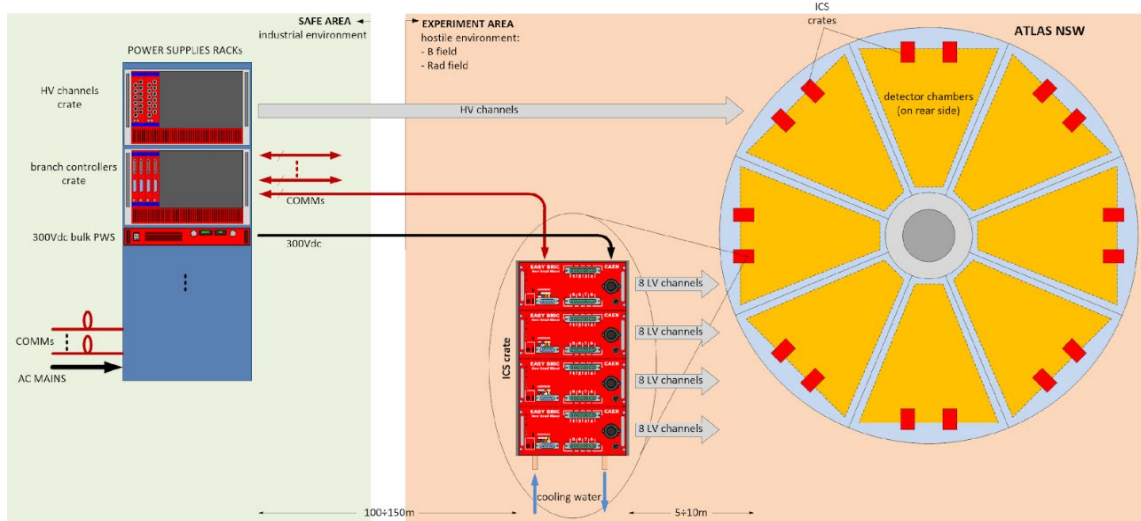

Figure 1: Power distribution strategy with the new DC/DC converter integrated in the detector.

# 2. The case of the ATLAS New Small Wheel

The ATLAS New Small Wheel (NSW) [1] upgrade is a good example for the case shown in Figure 1. It is composed of two detectors, the Micromegas (MM) for tracking and the sTGC for triggering. The low voltage power required by the front-end electronics of both detectors is about 100 kW. In case of installation of the low voltage power system in the balconies, considering an average distance to the front-end boards of 30 m and the copper cross-section allowed by its output connectors, 1.3 mm$^2$, the voltage drop on the cables at the nominal current of 16 A would have been about 12 V, making this solution impossible to pursue. Therefore, the final specifications [2, 3] of the system led to the choice of requesting on-wheel converters. The tender was awarded to the company CAEN S.p.A., which designed and produced the Easy 6000 BRIC [4]. The requirements for the design are summarized in Table 1.

| Electrical | | Environmental | |
|---|---|---|---|
| $V_{in}$ | 280 - 300 V DC | Ionising Radiation | 96 Gy |
| $V_{out}$ | +10 V DC ±10% | Displacement Damage | 5.8 x $10^{12}$ 1-MeV Eq. n/cm$^2$ |
| $I_{out}$ nominal | 16 A DC | Single-Event Fluence | 1.0 x $10^{12}$ p/cm2 (E > 20 MeV) |
| Efficiency | ≥ 80% above 50% load | Magnetic Field | 0.5 T |
| Ground | Floating, each ch. isolated | | |
| Com. ripple | < 50 mVpp 0 – 30 MHz | | |
| Diff. ripple | < 20 mVpp 0 – 30 MHz | | |

Table 1: Electrical and environmental requirements for NSW low voltage power supplies.

## 3. Design

Bringing DC/DC converters closer or inside the detectors opens for new technology challenges such as radiation and magnetic field tolerance. The increase of the input voltage of the converter sets new constraints in the choice of the electronic devices that will have to switch the current in the presence of a high radiation background. GaN devices have proven to be effective and were chosen for the presented design, as they are less sensitive to ionising radiation, working with higher voltages, currents and frequencies. Another criticality of the presented DC/DC is the requirement of designing an isolated converter, inserting a transformer, which need to work in the presence of a high magnetic field (up to 0.5 T in the case of NSW). Normal ferrite cores used in conventional switching electronics suffer from saturation well below 0.1 T: the mitigation of the problem was found in the choice of particular materials with low permeability for the core. A proper type of converter topology for the given needs is the LLC Half-Bridge converter, belonging to the resonant family (Figure 2). This type of circuit offers several advantages, low inductive component count, voltage applied to FETs defined and limited to Vin, fixed 50% duty-cycle driving and high power density. Key element is a resonant circuit based on an L-C tank, where L and C are both physical devices and the parasitic components of other devices in the circuit. By properly choosing the working point of the circuit, it is possible to achieve zero voltage switching (ZVS) at the primary side of the transformer; similarly, on the secondary, it is possible to let rectifying diodes open when the current passing through them approaches zero.

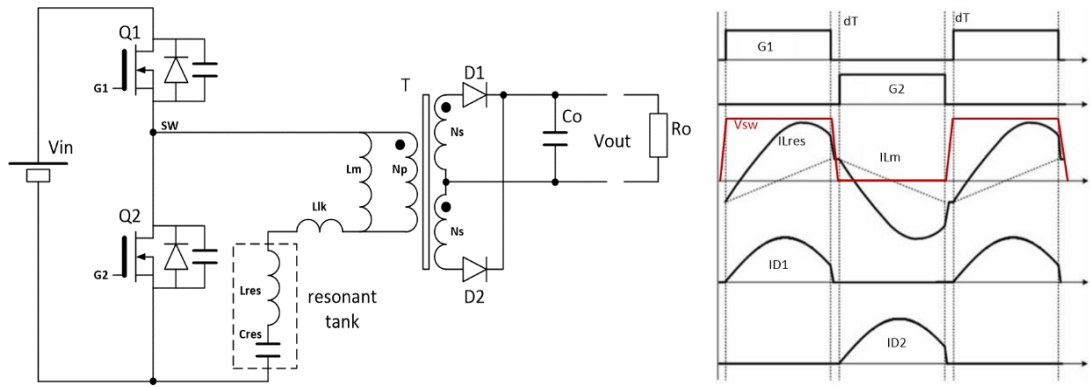

Figure 2: LLC Half-Bridge converter topology (left) and time diagram (right).

Soft switching reduces ringing, noise propagation and power dissipation. The regulation of the converter output takes place by modulating the switching frequency: LLC controller driving the FETs can vary the frequency in the range of 140 to 500 kHz (max to min power) to regulate the output voltage. To meet the NSW technical specifications, a module containing eight channels was designed (Figure 3 left): every channel is independently controlled and remotely monitored. A local controller manages the eight power channels and holds communications with a remote controller. Communications are carried out via RS-485 differential lines and are locally isolated on the module: an auxiliary power supply provides all the service voltages necessary for the circuits. Primary power for the auxiliary supply is 48 V, provided by the remote controller to be fully independent from the 300 V. Both local microcontroller and the isolation of communications lines were affected by choices of reliability in radiation environment: various preventive tests were carried out to choose the specific devices.

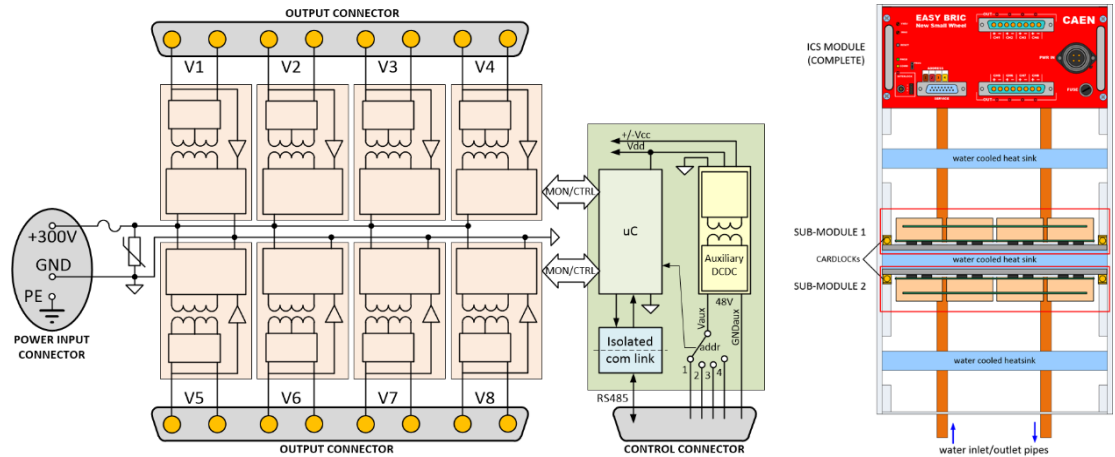

Figure 3: Internal schematic block of a converter module (left) and modules mechanical arrangement in the water-cooled crate (right).

Another requirement was grouping four modules in a unique crate (Figure 3 right), therefore the power channels are divided into two submodules of four channels each. Mechanical engineering of the converter modules and crate was highly dependent on heat management: since the presence of a magnetic field prevented the use of fans, the heat transfer produced by the module was achieved by a water cooling system. All components that need to dissipate heat are brought into thermal contact with an aluminum plate via a thermal pad at its time pressed on heatsinks with cardlocks. Functional cards such the ones hosting the microcontroller, the auxiliary power supply, the communications, the service connector and the power input and outputs are common for a module and are located in a front panel. This has a back card that connects all the functional cards together and to the submodules, closing the front part of the crate and allowing quick maintenance operations.

## 4. Environmental tests

The produced prototypes were tested in facilities available at CERN to qualify them for acceptance and production. Concerning radiation tests, three campaigns were performed in CHARM [5] on the prototypes to converge to the final design. At the end of the last irradiation, the prototype collected the following doses/fluences: Total Ionizing Dose (TID) = 189.6 Gy; Hadrons (E > 20 MeV) = 6.7 x $10^{11}$/cm$^2$; Neutrons (1 MeV equivalent) = 2.1 x $10^{12}$/cm$^2$.

All channels were still working after the irradiation, microcontroller internal memory kept the settings and calibration data without any corruption and internal auxiliary voltages, channels output voltages/currents monitor remained constant and stable. The collected events were 30: three of them were "OFF", meaning that the channels went OFF and required a manual intervention to be put again in ON state; seven of them were "communications lost", and again required a manual reset to restore them. The remaining 20 were self-resets of the microcontroller.

Concerning magnetic field tests, a behavioural hardness and efficiency measurement was done at CERN with the magnet MNP17: all channels were connected to active loads and switched ON at different magnetic field, from 0 up to 0.5 T in steps of 0.1 T, and at different output currents, up to 15 A (Figure 4). A second magnetic field test was done in the North Area at the GOLIATH facility [6], to check the performances at extreme condition (up to 1.3 T) by looking voltage/current waveforms at the switching node in the primary side of the transformer [7].

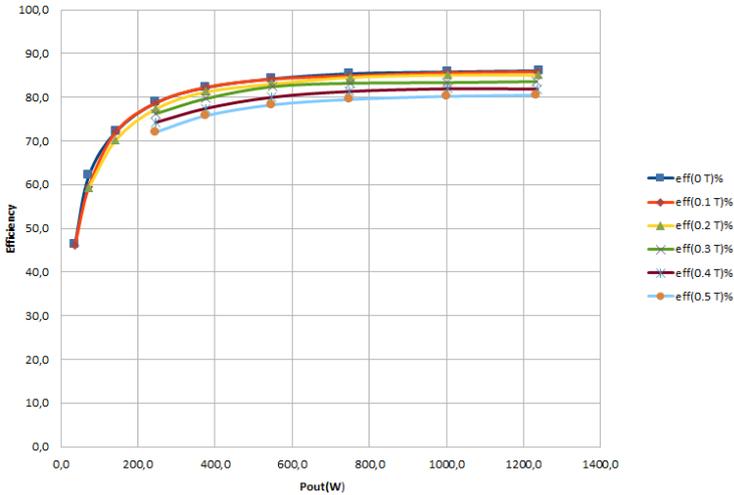

Figure 4: efficiency vs. output power.

## 5. Performances and optimization

A test setup has been organised in order to measure the main electrical performance of the converter. Measured parameters were:

- Line regulation as a function of the input voltage and the output current (Figure 5): output voltage was measured, reducing progressively the primary input voltage. The measurement was repeated for different constant loads (9 A, 12 A and 15 A).
- Load regulation as a function of the output current (Figure 6): output voltage was measured, increasing progressively the load current.
- Load step response (Figure 7: output current in yellow trace and AC-coupled output voltage in green trace): a constant 8 A load was stepped to 12 A and sequentially reduced again to 8 A in 80 ms. Output voltage was fixed to 10.5 V.
- Output differential ripple, with AC-coupled differential probe (Figure 8).

As isolated power supply for detectors, one of the most critical parameter is the common mode noise. All precautions in the matter of filtering were taken to reduce the noise propagation up to the front-end electronics. Two refurbishment campaigns were done to converge progressively to the optimum readout noise level: the first campaign was done during the commissioning of the single detectors and the second when the system included all the converters, cabling and all the detectors were installed on the wheels. Due to the presence of the external magnetic field, the only mode of intervention was inserting capacitors in specific points of the circuit. Figure 9 and Figure 10 represents the output common mode noise before and after the global refurbishment: an overall reduction of a factor 4 to the peak-to-peak noise has been achieved.

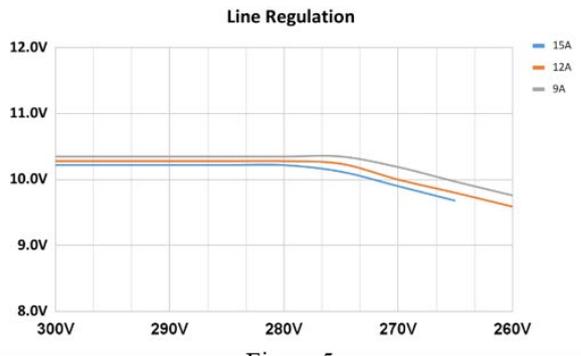
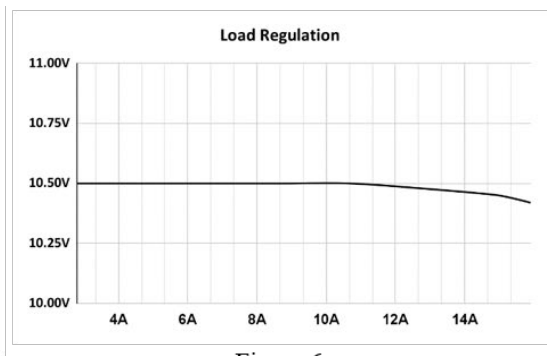

Figure 5

Figure 6

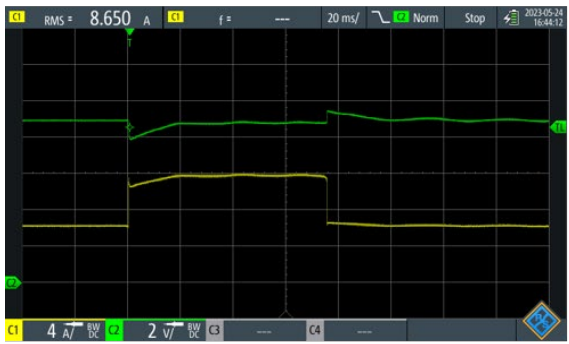
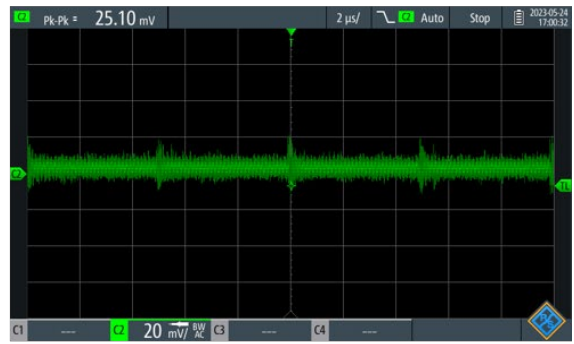

Figure 7: step response.

Figure 8: diff. mode ripple.

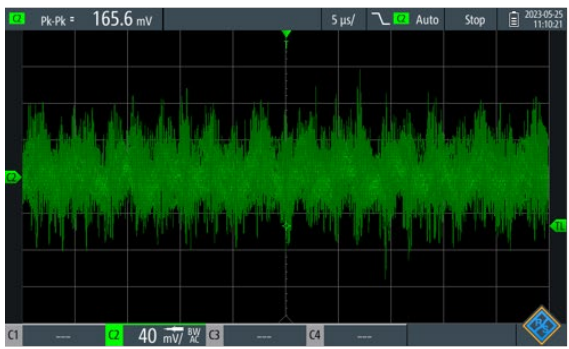
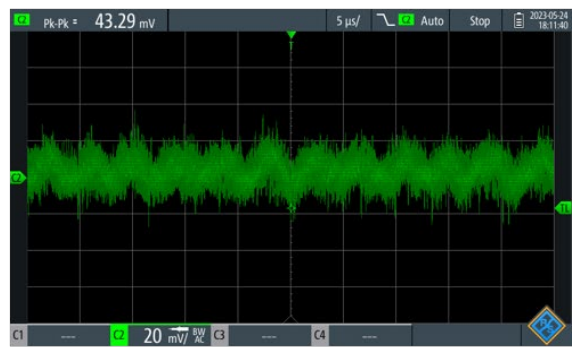

Figure 9: comm. mode noise, before optimization.

Figure 10: comm. mode noise, after optimization.

## 6. Conclusions

A new on-detector-installed system with high power density able to supply the front-end electronics in large experiments has been described. It works in a harsh radiation environment and in presence of high magnetic field. Electrical performances are shown, together with the results of the validation in a hostile environment.

## Acknowledgments

The authors are indebted to their colleagues that participated and gave a precious contribution to the various tests and validations: D. Calabrò (INFN Pavia), G. Di Maio and M. Rogai (CAEN S.p.A.), G. Giunta and M. Pirola (University and INFN Pavia). We would like to thank C. Paraskevopoulos (LNF of INFN) for automatizing the burn-in tests, I. Mesolongitis (University of West Attica) for the design of a common mode filter and A. Toro Salas (Brandeis University) for the EMI/EMC measurements of the prototypes.